\documentclass{iau}
\usepackage{graphicx}
\usepackage{natbib}
\usepackage{aas_macros}
\usepackage{multicol}

\title[The Physics of Galaxy Evolution with SPICA] 
{Unveiling the physical processes that regulate Galaxy Evolution with SPICA observations}

\author[Spinoglio, Fern\'andez-Ontiveros \& Mordini]   
{Luigi Spinoglio, 
 Juan A. Fern\'andez-Ontiveros \and Sabrina Mordini}

\affiliation{
Istituto di Astrofisica e Planetologia Spaziali - INAF, Rome, \\ Via Fosso del Cavaliere 100, 00133, Roma, Italia\\
email: {\tt luigi.spinoglio@iaps.inaf.it, j.a.fernandez.ontiveros@gmail.com, sabrina.mordini@uniroma1.it} \\[\affilskip]
}

\pubyear{2019}
\volume{356}  
\jname{Nuclear Activity in Galaxies Across Cosmic Time}
\editors{P. Marziani, J. Masegosa, S.H. Negu, H. Netzer, M. Povic, P. Shastri {\&} S.B. Tessema, eds.}

\begin{document}

\maketitle

\begin{abstract}
To study the dust obscured phase of the galaxy evolution during the peak of the Star Formation Rate (SFR) and the Black Hole Accretion Rate (BHAR) density functions ($z = 1 - 4$), 
rest frame mid-to-far infrared (IR) spectroscopy is needed. At these frequencies, dust extinction is at its minimum and a variety of atomic and molecular transitions, tracing most astrophysical domains, occur. The future IR space telescope mission, {\it SPICA}, fully redesigned with its $2.5\, \rm{m}$ mirror cooled down to $T < 8\, \rm{K}$, will be able to perform such observations. With {\it SPICA}, we will: 1) obtain a direct spectroscopic measurement of the SFR and of the BHAR histories, 2) measure the evolution of metals and dust to establish the matter cycle in galaxies, 3) uncover the feedback and feeding mechanisms in large samples of distant galaxies, either AGN- or starburst-dominated, reaching lookback times of nearly 12 Gyr. {\it SPICA} large-area deep surveys will provide low-resolution, mid-IR spectra and continuum fluxes for unbiased samples of tens of thousands of galaxies, and even the potential to uncover the youngest, most luminous galaxies in the first few hundred million years. In this talk a brief review of the scientific preparatory work that has been done in extragalactic astronomy by the {\it SPICA} Consortium will be given.

\keywords{telescopes, galaxies: evolution, galaxies: active, galaxies: starburst, quasars: emission lines, galaxies: ISM, galaxies: abundances, galaxies: high-redshift, infrared: galaxies}
\end{abstract}

\firstsection 
\section{Introduction}
\begin{figure}[t]
\begin{center}
    \includegraphics[width=12cm]{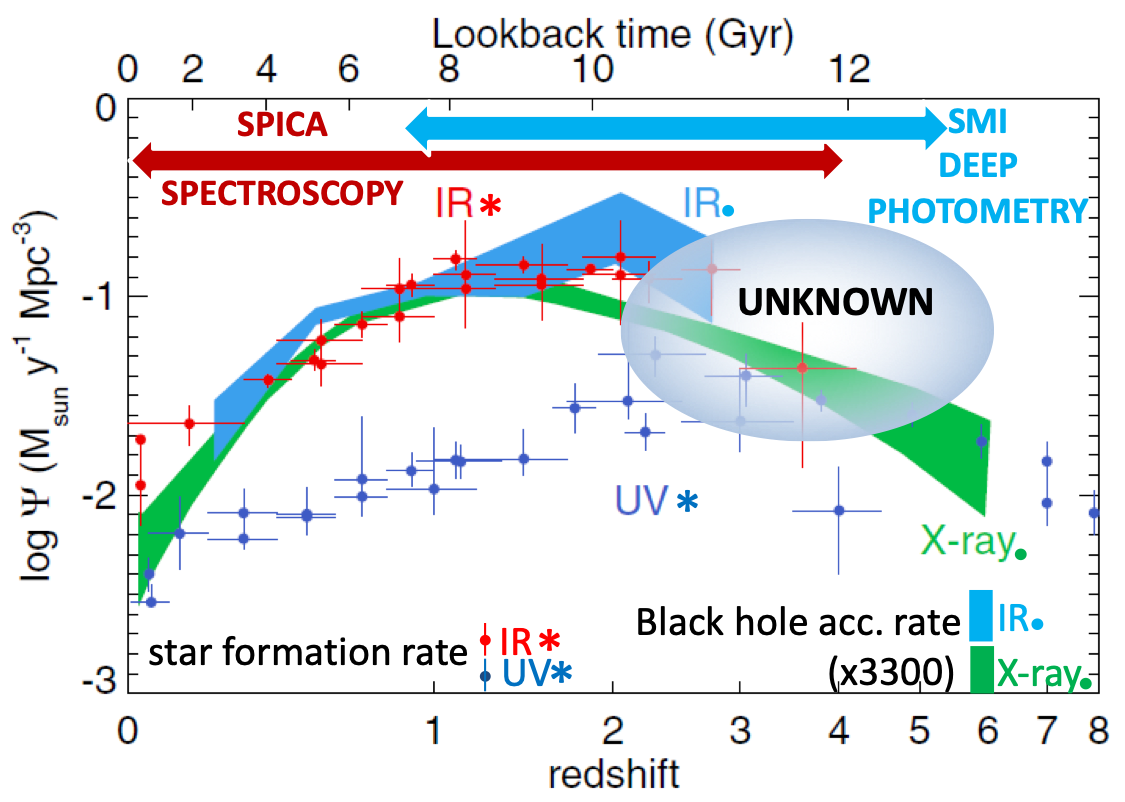}
    \caption{Estimated star-formation rate densities from the far-ultraviolet (blue points) and far-IR (red points) photometric surveys \citep[figure adapted from][]{ma_di14}. The estimated BHAR density, scaled up by a factor of $3300$, is shown for comparison (in green shading from X-rays and light blue from the IR).}\label{fig_madau}
\end{center}
\end{figure}

The bulk of the star formation and supermassive black hole (SMBH) accretion in galaxies took place more than six billion years ago, with a sharp drop to the present epoch \citep[e.g.,][Fig.\,\ref{fig_madau}]{ma_di14}. Since most of the energy emitted by stars and accreting SMBHs is absorbed and re-emitted by dust, understanding the physics of galaxy evolution requires infrared (IR) observations of large, unbiased galaxy samples spanning a range in luminosity, redshift, environment, and nuclear activity. From {\it Spitzer} and {\it Herschel} photometric surveys the Star Formation Rate (SFR) and Black Hole Accretion Rate (BHAR) density functions have been {\it estimated} through measurements of the bolometric luminosities of galaxies. However, such integrated measurements could not separate the contribution due to star formation from that due to BH accretion. This crucial separation has been attempted so far through modelling of the spectral energy distributions and relied on model-dependent assumptions and local templates, with large uncertainty and degeneracy. On the other hand, determinations from UV and optical spectroscopy (e.g. the SLOAN survey) track only marginally ($\sim$10\%) the total integrated light (Fig.\,\ref{fig_madau}). X-ray analyses of the BHAR, in turn, are based on large extrapolations and possibly miss a large fraction of obscured objects.
Furthermore, the SFR density at $z > 2 - 3$ is very uncertain, since it is derived from UV surveys, highly affected by dust extinction. As opposite, through IR emission lines, the contributions from stars and gravity can be separated. {\it SPICA} spectroscopy will allow us to directly measure redshifts, SFRs, BHARs, metallicities and dynamical properties of gas and dust in galaxies at lookback times up to about 12 Gyrs. {\it SPICA} spectroscopic observations will allow us for the first time to redraw the SFR rate and BHAR functions (Fig.\,\ref{fig_madau}) in terms of measurements directly linked to the physical properties of the galaxies. 
\vspace*{-0.6 cm}
\section{Why IR spectroscopy}

\begin{figure}[t]
    \includegraphics[width=5.2cm]{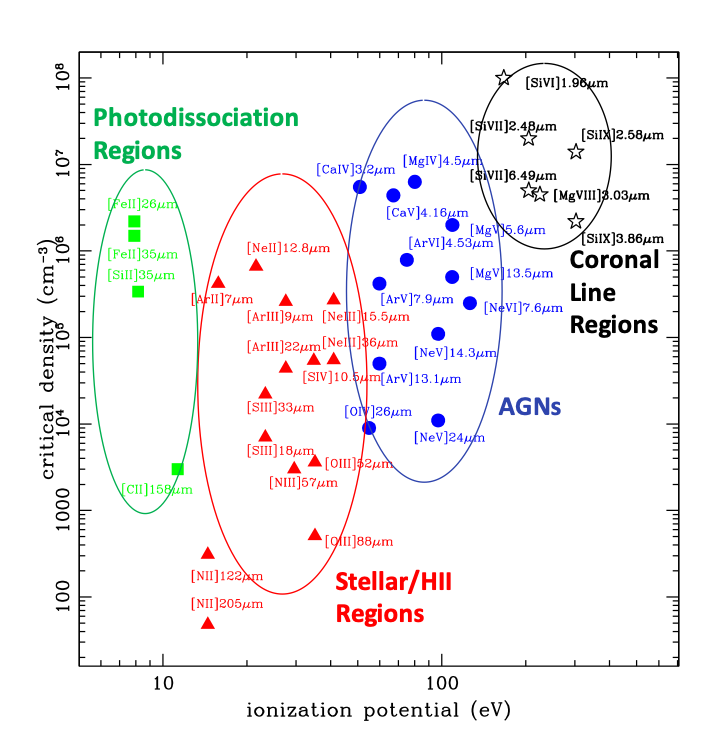}~
    \includegraphics[width=8.8cm]{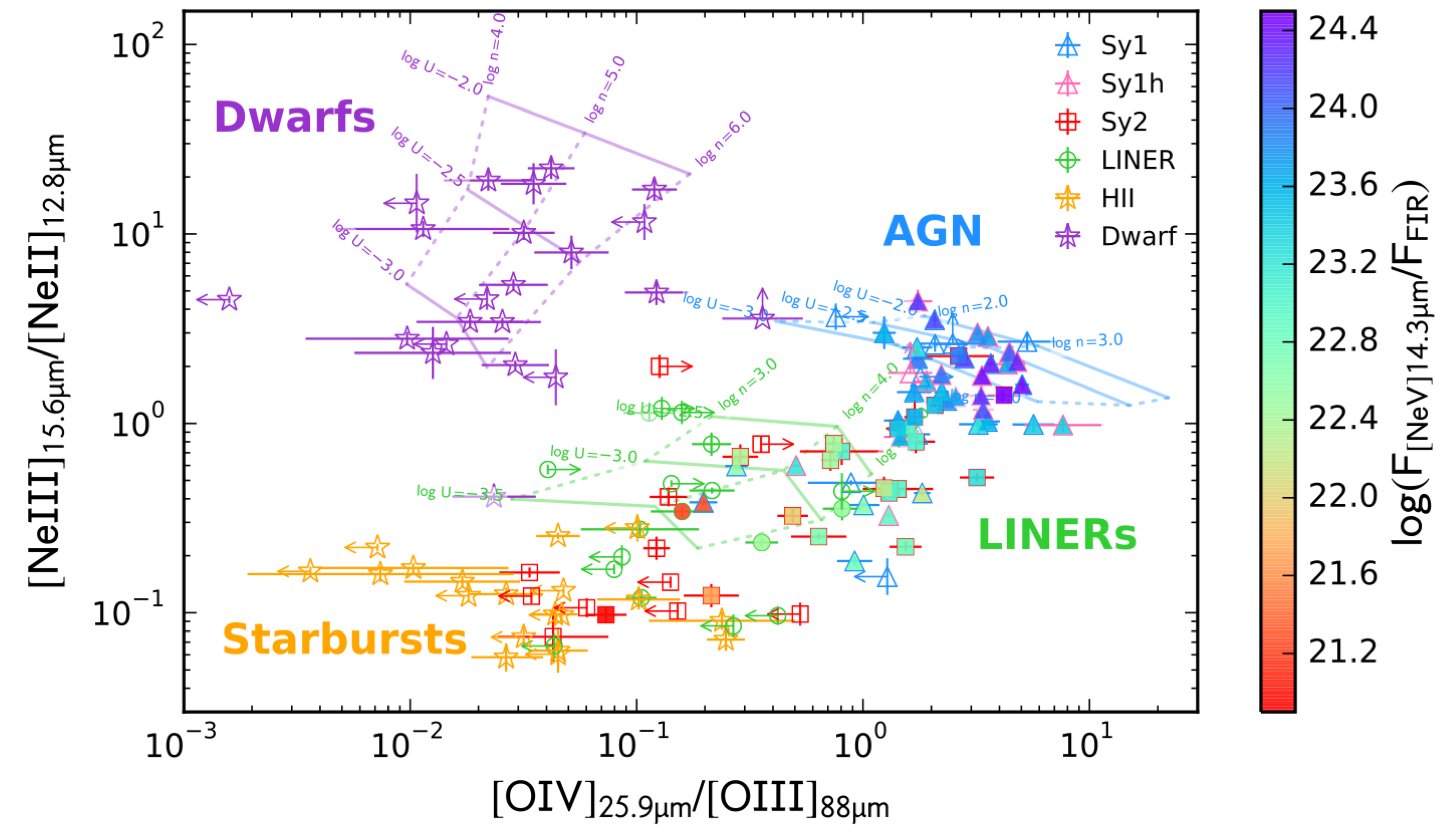}
    \caption{{\bf Left:} the ionization density diagram of the IR fine structure lines \citep{sm92}. {\bf Right:} Observed line ratios of [NeIII]15.6$\mu$m/[NeII]12.8$\mu$m vs. [OIV]26$\mu$m/[OIII]88$\mu$m for AGN, LINER, starburst and dwarf galaxies in the local Universe. Figure adapted from \citet{fer16}.}\label{fig_IR_BPT}
   \vspace*{-0.4cm}

\end{figure}

The mid- to far-IR spectral range hosts a suite of atomic and ionic transitions, covering a wide range of excitation, density, and metallicity, directly tracing the physical conditions in galaxies, which are typically obscured during most of their evolution. Ionic fine structure lines (e.g. [NeII], [SIII], [OIII]) probe HII regions around hot young stars, providing a measure of the SFR and the gas density. Lines from highly ionized species (e.g. [OIV], [NeV]) trace the presence of AGN and can measure the BHAR. Photo-dissociation regions (PDR), the transition between young stars and their parent molecular clouds, can be studied via the strong [CII] and [OI] lines \citep[][Fig.\,\ref{fig_IR_BPT}-a]{sm92}. 

Through line ratio diagrams, like the {\it new IR BPT diagram} \citep[][Fig.\,\ref{fig_IR_BPT}-b]{fer16}, IR spectroscopy can separate the galaxies in terms of both the source of ionization --\,either young stars or AGN excitation\,-- and the gas metallicity, during the dust-obscured era of galaxy evolution ($0.5 < z < 4$). 


\vspace*{-0.6 cm}

\section{SPICA observations of galaxy evolution}

The SPace Infrared telescope for Cosmology and Astrophysics ({\it SPICA}) will combine a $2.5\,\rm{m}$ mirror cooled to below $8\, \rm{K}$ with instruments employing state-of-the-art detectors, becoming the first large telescope cooled in space using mechanical cryo-coolers instead of liquid cryogen. The low telescope background and the new generation of detectors will provide about two orders of magnitude sensitivity improvement with respect to previous missions. {\it SPICA} instruments will provide a spectral resolving power ranging from $R = 50 - 120$ to $11000$ in the $17 - 230\, \rm{\mu m}$ domain as well as $R \sim 28000$ between $12 - 18\, \rm{\mu m}$. The Transition Edge Superconductor detectors in the SAFARI spectrometer ($35 - 230\, \rm{\mu m}$) will reach an unprecedented sensitivity of $\sim 7 \times 10^{-20}\, \rm{W\,m^{-2}}$ ($5 \sigma$/1\,hr). Thanks to its large field of view of $10' \times 12'$, the SMI imager and spectrometer will deliver simultaneous spectroscopy ($17 - 37\, \rm{\mu m}$) and photometry ($34\, \rm{\mu m}$) mapping for large areas in the sky either. Additionally the B-BOP instrument will provide accurate polarimetric imaging at $70$, $220$ and $350\, \rm{\mu m}$ \citep{and19}. A full description of the mission, the telescope and the focal plane instruments can be found in \citet{roe18}.

In the field of galaxy evolution, {\it SPICA} will: 1) obtain the first spectroscopic characterization of the SFR and the BHAR histories \citep{spi17};
2) measure the evolution of metals and dust and establish the matter cycle in galaxies \citep{fer17}; 3) uncover the feedback and feeding mechanisms in large samples of distant galaxies, either AGN- or starburst-dominated \citep{gon17}; 4) provide low-resolution, mid-IR spectra and continuum fluxes for deep unbiased samples of tens of thousands of galaxies, and even the potential to uncover the youngest, most luminous galaxies in the first few hundred million years \citep{gru17, kan17}; 5) probe the spectra of hyper-luminous IR galaxies at redshift $z = 5 - 10$, allowing us to characterize their main physical properties \citep{ega18}.

\subsection{Feedback from powerful AGN}

The correlations between the SMBH mass and the velocity dispersion, stellar mass, and luminosity of galaxies in the local Universe \citep{mag98, fer00} suggests a link between the growth of the BH and the stellar population in its host galaxy. 
The bimodal color distribution observed in local galaxies \citep{str01, bal04} points to an scenario where massive red-and-dead galaxies finished their evolution on very short timescales, while the evolution of low-mass blue galaxies is still ongoing \citep{hop06, sch14}. But even more relevant for our current (poor) knowledge of galaxy evolution would be to reconcile the shape of the observed luminosity function of galaxies with that of the theoretical halo mass function in CDM. Feedback from Supernovae can match observations with theory at low masses, while AGN feedback would be needed to make them agree in the high mass end. In this self-regulated feedback model, the funneling of large amounts of gas into the nuclear region generates both a nuclear starburst (SB) and drives the accretion onto the SMBH. The latter eventually reaches a critical mass/luminosity when the energy and momentum released couples with the surrounding interstellar medium (ISM), limiting the accretion onto the SMBH and quenching the SBs via injection of turbulence, through a fast sweeping out of the ISM gas reservoir, or by heating the circumgalactic gas and preventing further accretion of gas onto the galaxy (negative feedback), ultimately yielding the MBH-$\sigma$ relationship \citep{sil98,dim05,spr05,mur05,hop06}.

\begin{figure}[b]
\begin{center}
    \includegraphics[width=6cm]{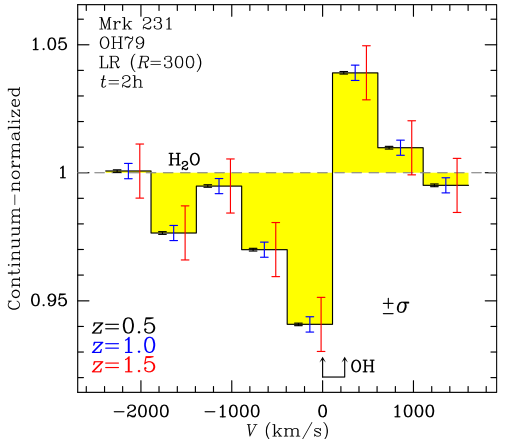}~
    \includegraphics[width=6cm]{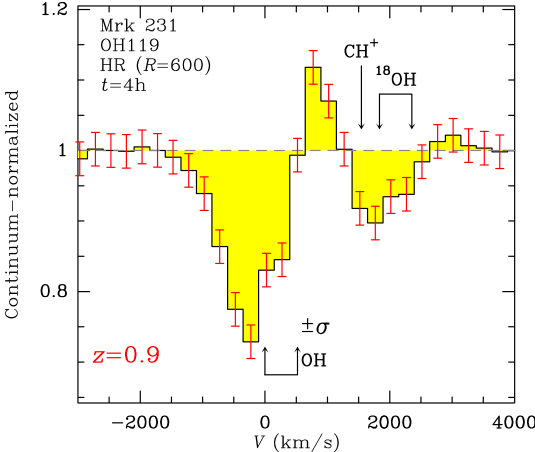}
    \caption{{\bf Left:} The OH 79$\mu$m P Cygni profile observed by {\it Herschel} in Mrk231 \citep{fis10} and simulated with the SAFARI low resolution mode at z=0.5, 1.0, 1.5 (see errorbars) \citep{gon17}. {\bf Right:} Same, but simulated with the SAFARI FTS high resolution mode at redshit of z=0.9 \citep{gon17}.}\label{fig_out}
\end{center}
\end{figure}

With {\it SPICA} we will be able to address the following questions: is AGN feedback responsible for the decline of SF in the last 7 Gyr, driving massive galaxies into the red-and-dead sequence? What physical processes --\,mechanical or thermal energy injection\,-- drive molecular outflows? {\it SPICA} will be able to detects P-Cygni profiles in the OH lines of powerful ultra-luminous galaxies up to a redshift of $z \sim 2$ in low resolution mode, and up to $z \sim 1$ at high spectral resolution, as shown in Fig.\,\ref{fig_out}. 

\subsection{Measuring metallicity evolution with {\it SPICA}}

To study the chemical evolution of galaxies, especially during the dust-obscured era across the peak of the SFR density ($1 < z < 3$), the use of metallicity tracers almost independent of the dust extinction, radiation field, and of the gas density are crucial. Optical nebular lines are likely probing just the most external (unobscured) regions of these galaxies, as suggested by the order of magnitude discrepancy found between the dust content of high redshift sub-millimetre galaxies (SMG) seen by {\it Herschel} and the metallicity measurements based on optical lines \citep{san10}. This is likely caused by dust obscuration, since the ISM of SMG is optically thick at visual wavelengths, thus the optical emission in these galaxies comes probably from the outer parts which would be poorly enriched with heavy elements.

\begin{figure}[t]
\begin{center}
    \includegraphics[width=6cm]{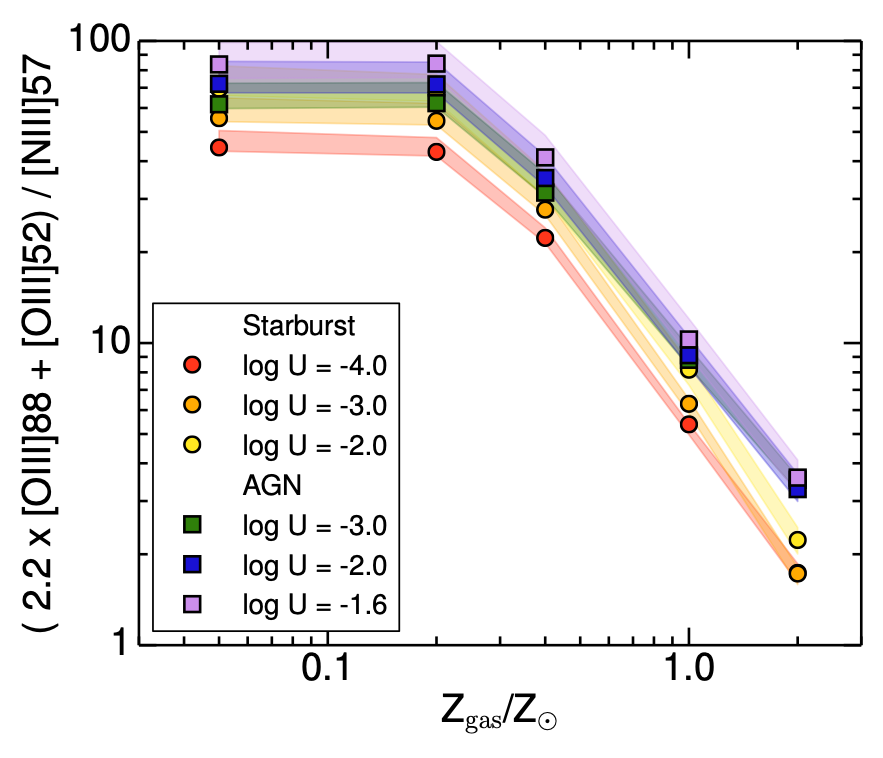}~
    \includegraphics[width=7.3cm]{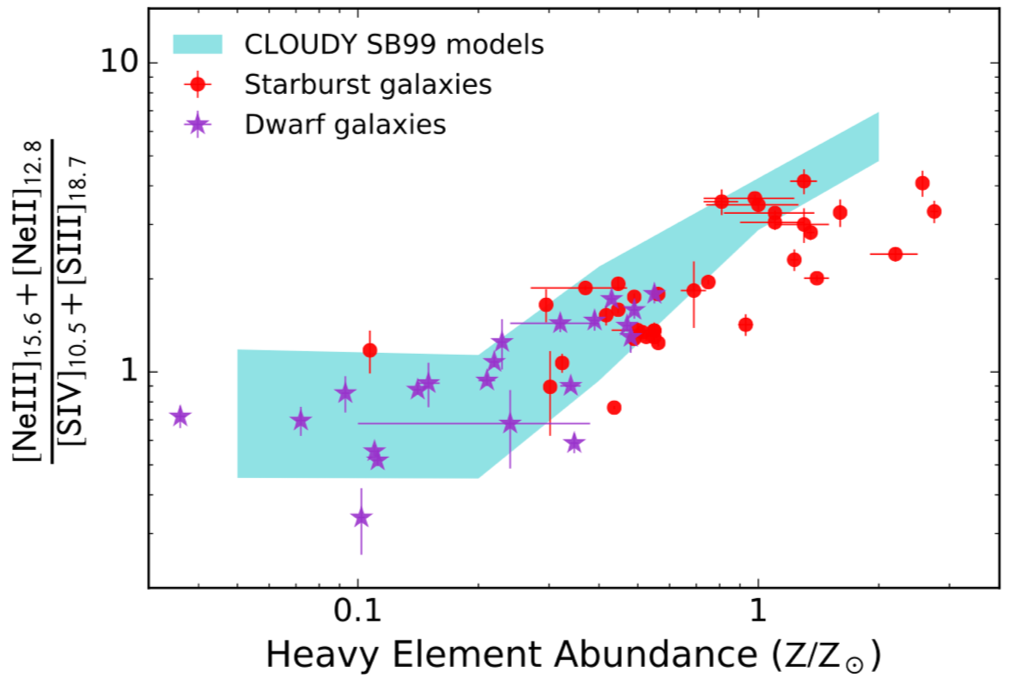}
    \caption{{\bf Left:} AGN and starburst models for the metallicity sensitive ($2.2 \times$\,[OIII]88$\mu$m+[OIII]52$\mu$m)/[NIII]57$\mu$m line ratio as a function of the gas-phase metallicity \citep{per17}.
    {\bf Right}: The ([NeII]12.8$\mu$m + [NeIII]15.6$\mu$m) to ([SIII]18.7$\mu$m + [SIV]10.5$\mu$m) line ratio from Spitzer/IRS observations of local starburst galaxies vs. indirect gas-phase metallicity determined from strong optical lines (Moustakas et al. 2010; Pilyugin et al. 2014). Cloudy simulations including sulphur stagnation above $Z > 1/5 Z_{\odot}$ are in agreement with the observed increase of this line ratio \citep{fer16,fer17}.} \label{fig_met}
\end{center}
\end{figure}

Gas metallicities can be determined in dust-obscured regions and galaxies using the diagnostic shown in Fig.\,\ref{fig_met}-a, up to $z \sim 1.6$, where the [OIII]88$\mu$m line would still fall within the spectral range of SAFARI. A diagnostic based on the [OIII]52$\mu$m/[NIII]57$\mu$m ratio (see Nagao et al. 2011) is also a metallicity tracer up to z$\sim$3 if the density is constrained through other line ratios. Above $z > 0.15$ and $0.7$, the [SIV]10.5$\mu$m line would enter in the SMI/HR and MR ranges, respectively, enabling an additional indirect abundance diagnostic --\,based on the calibration of metallicity-sensitive line ratios\,-- using the ([NeII]12.8$\mu$m + [NeIII]15.6$\mu$m) to ([SIII]18.7$\mu$m + [SIV]10.5$\mu$m) line ratio (Fig.\,\ref{fig_met}-b; \citealt{fer16}). 

\vspace{0.2cm}

\section*{Acknowledgements}
\textit{We acknowledge the whole SPICA Collaboration Team, as without his multi-year efforts and work this paper could not have been possible. LS and JAFO acknowledge financial support by the Agenzia Spaziale Italiana (ASI) under the research contract 2018-31-HH.0.}
\vspace*{-0.2cm}

{\small
\begin{multicols}{2}

\end{multicols}
}


\end{document}